\documentstyle[12pt,aasms4,epsfig]{article}
\received{9 July 1997}
\accepted{3 October 1997}
\input{psfig}
\lefthead{Yu et al.}
\righthead{Kilo-hertz QPO in Island State of 4U1608-52 as Observed with
RXTE/PCA}
\begin{document}
\title{Kilo-hertz QPO in Island State of 4U1608-52 as Observed with RXTE/PCA}
\author{W. Yu\altaffilmark{1}, S. N. Zhang\altaffilmark{2,3}, 
B .A. Harmon\altaffilmark{3}, W. S. Paciesas\altaffilmark{3,7}, 
C. R. Robinson\altaffilmark{2,3}, J. E. Grindlay\altaffilmark{4}, 
P. Bloser\altaffilmark{4}, D. Barret\altaffilmark{5}, 
E. C. Ford\altaffilmark{6}, M. Tavani\altaffilmark{6}, P. 
Kaaret\altaffilmark{6}}
\altaffiltext{1}{LCRHEA, Institute of High Energy Physics, Beijing 100039, China 
} 
\altaffiltext{2}{ Universities Space Research Association, Huntsville, AL 35806 
} 
\altaffiltext{3}{ NASA/Marshall Space Flight Center, ES-84, Huntsville, AL 35812 
} 
\altaffiltext{4}{ Harvard-Smithsonian Center for Astrophysics, 60 Garden Street,
Cambridge, MA 02138 } 
\altaffiltext{5}{ Centre d'Etude Spatiale des Rayonnements, CNRS-UPS, BP4346, 
31028 Toulouse Cedex 04, France } 
\altaffiltext{6}{Columbia Astrophysics Laboratory, Columbia University, 538
West 120th Street, New York, NY 10027 } 
\altaffiltext{7}{ University of Alabama in Huntsville, Department of Physics, 
Huntsville, AL 35899 } 

\begin{abstract} 

We report RXTE/PCA observations of 4U 1608-52 on March 15, 18 and 22 
immediately after the outburst in early 1996. The persistent count rates 
ranged from 190 to 450 cps (1-60 keV). During this period of time, 4U 1608-52 
was in the island state. We detected QPO features in the power density spectra 
(PDS) at 567-800 Hz on March 15 and 22, with source fractional root-mean-square 
({\it rms}) amplitude of 13\%-17\% and widths of 78-180 Hz. The average 
{\it rms} amplitude of these QPO features is positively correlated with the 
energy. Our results imply that the neutron star spin frequency is possibly 
between 300 Hz and 365 Hz. 

\end{abstract}

\keywords{X-ray:stars --- stars:individual(4U 1608-52) --- stars: neutron}

\section{Introduction}
4U1608-52 is a transient low-mass X-ray binary with outbursts which 
recur on timescales of 80 days to two years (\cite{grindlay78};
\cite{lewin93}; \cite{lochner94}). 
It was classified to be an atoll source based on the correlated X-ray spectral
variability and high-frequency-noise (HFN) in the X-ray intensity (\cite{HK89};
\cite{yoshida93}). 4U 1608-52 has been observed a few times with an energy
spectrum consistent with a power-law at X-ray luminosities below 
${10}^{37}$ergs/s (\cite{mitsuda89}; \cite{penninx89}; \cite{yoshida93}; 
\cite{zhang96}).

The X-ray monitoring of 4U1608-52 with RXTE/ASM indicated an 
outburst in early 1996. In response to a high state detection of 4U1608-52 
in RXTE/PCA scans (\cite{Marshall96}), pointed observations with the RXTE/PCA 
were conducted during the decay phase of this outburst on March 3, 6, 9, and 12 
(\cite{berger96}; \cite{klis97a}), and on March 15, 18, and 22 (see Fig.1). 
Kilo-hertz QPOs were discovered in 4U 1608-52 in March 3, 6, and 9 observations. 
No X-ray bursts were observed with RXTE/PCA in early March 1996 
(\cite{berger96}). Here we report the timing analysis results on RXTE/PCA 
observations of 4U 1608-52 on March 15, 18, and 22. 

\section{Observations and Analysis Results}

X-ray monitoring by RXTE/ASM shows that our observations on March 15, 18 and
22 were taken near the end of the outburst decay, with ASM daily-averaged 
brightness 
of $44.8\pm5.2$, $13.3\pm7.2$, and $14.5\pm4.1$ mCrab respectively (see Fig.1). 
The 
persistent X-ray flux (2-20 keV) obtained with the RXTE/PCA was in the range 
between
$4.6\times{10}^{-10}$erg/s/cm$^{2}$ and 
$1.1\times{10}^{-9}$erg/s/cm$^{2}$. The source count rate in the energy range 
1-60 keV varied between 190 and 450 cps. Three X-ray bursts were observed 
in one orbit on March 22. They show the evidence of a high energy excess above a 
Planckian spectrum, but no kilo-hertz QPO were detected in the bursts (Yu et al. 1997, 
in preparation). We thus exclude 50 seconds data of each burst when estimating 
the properties of the kilo-hertz features in the persistent emission.  

The PCA/RXTE provides several data modes which were used in the 
analysis reported in this paper. The color-color diagrams were constructed from 
the {\it Standard 2 Mode} data, which provides 129 energy channels and 16 second 
time resolution. We made use of the {\it Event Mode} data (64 energy channels 
and 122 $\mu$s time resolution) to generate the power density spectra (PDS) 
from 0.001 Hz to 100 Hz.
\subsection{Color-Color Diagrams}
Background count rates as a function of time were produced with the standard 
background model supplied by the RXTE Guest Observer Facility (\cite{stark97}). 
After subtracting the background in 3 bands (approximately 2.2-5.1, 5.1-10.1, 
and 10.1-29.8 keV), light curves and color-color diagrams for 4U1608-52 were 
generated. Fig.2 is the color-color diagram of 4U 1608-52 plotted with the data 
obtained on March 15, 18, and 22, as triangles, squares, and circles, respectively. 
Each data point represents 240s of observation. The data for  about 2000s, starting 
from the rise of the first burst to the end of that orbit, were excluded. The 
uncertainty in the hardness ratios were also estimated for a hypothetical EXOSAT 
observation, as shown in Fig.2.  
\subsection{Power Density Spectra (PDS)}
 We have obtained rebinned PDS in the frequency range of ${10}^{-3}-100$ Hz 
 from background-subtracted light curve of the {\it Event Mode} data. The average 
 PDS of individual days on March 15, 18 and 22 are shown in Fig. 3. The average 
 level of the white noise caused by counting statistics was subtracted in these 
 spectra. These PDS show similar HFN components with a flat top from 0.01 Hz to 
 about 10 Hz. The low frequency noise (LFN) can be represented by a power-law and 
 the HFN components can be described as a power-law with an exponential cutoff 
 for atoll sources (\cite{HK89}). This model basically agrees with our data as 
 shown in Fig.3. The fractional {\it rms} of the HFN for each day is $12.0\pm1.9$\%, 
 $13.5\pm1.6$\%, and $14.4\pm0.6$\%, respectively. The correponding HFN cut-off 
 frequencies are $25\pm6$, $8\pm2$, and $19\pm3$ Hz. All the above errors represent 
 a 90\% confidence level (\cite{numerical92}).  

In the PDS of the first orbit on March 15, a QPO peak at 20 Hz can be seen. 
A broad excess of power at 10-30 Hz is also visible in the second and third 
orbits on March 15. Further analysis of the {\it Single-Bit Mode} data and the 
{\it Event Mode} data in each orbit at higher frequencies revealed QPO features 
in the frequency range of 567-800 Hz in the average PDS. An example of the 
Leahy-normalised PDS obtained from the {\it Event Mode} data in 1-30 keV in the 
3rd orbit on March 15 is shown in Fig. 4. There is broad excess of power between 
a few tens Hz to about 200 Hz in each of the PDS on March 15 and 22. The results 
are listed in Table 1. All errors in the table correspond to unreduced 
$\Delta{\chi}^{2}=1$. We estimate the {\it rms} amplitudes of the kilo-hertz QPO 
peaks by fitting PDS at 200-2000 Hz range with a linear component plus a Lorentzian 
peak. The {\it rms} amplitudes of the QPO peaks were obtained from the Lorentzian 
component. 

We then divided the 2-60 keV energy range into 7 bands ( 2.2-3.5 keV, 
3.5-5.4 keV, 5.4-7.9 keV, 7.9-11.1 keV, 11.1-14.6 keV, 14.6-20.4 keV and above 
20.4 keV) and studied the PDS in these bands using the {\it Event Mode} data. The 
QPO peaks were usually not clearly visible in the average PDS in each of the 7 bands. 
The individual PDS in the frequency range between 200 Hz and 2000 Hz were fit with a 
model composed of a linear component plus a box function. The box functions 
were centered at the peak frequencies and widths were set to be 2 times 
of the QPO FWHMs (see Table 1). Then we calculated the {\it rms} amplitude 
from the box function integrals. Finally we averaged the {\it rms} amplitudes 
from each energy-dependent PDS over the six orbits. A positive correlation between 
the average QPO {\it rms} amplitude and the photon energy was observed, as shown in 
Fig.5.

We also notice that there is no apparent correlation between the QPO {\it rms} 
and the intensity. But the QPO centroid frequencies and the average source intensities 
show the tendency of a positive correlation within a few hours (see Fig.6). For example, on March 15, 
the higher the count rates, the higher the QPO frequencies were. On March 22, a similar 
trend was also observed.
\section{Discussion}
Kilo-hertz QPOs in X-ray flux have been observed from about twelve LMXBs
so far (\cite{klis97b} and references therein; \cite{wzhang97}).
Nine of them are atoll sources or probable atoll sources.  
In our observation of 4U 1608-52, its persistent flux in the 2-20 keV
band ranged between $(4.6-11) \times {10}^{-10} \, {\rm erg} \, {\rm
s}^{-1} \, {\rm cm}^{-2}$, corresponding to an X-ray luminosity of
$(0.7-1.7) \times {10}^{36} \, {\rm erg} \, {\rm s}^{-1}$ assuming a
distance of 3.6 kpc (\cite{nakamura89}).  The luminosity was at or
below the lowest luminosity of 4U~1608-52 in its island state ever
observed with EXOSAT and Ginga (\cite{penninx89}; \cite{yoshida93}).
We detected no large motion in the PCA color-color diagram within a day.  
Also, the power density spectra were dominated by HFN with
{\it rms} around 13\%.  This suggests that 4U 1608-52 was in its
island state (\cite{HK89}; \cite{klis95}).  It is the second atoll
source observed to exhibit kilo-hertz QPO in both the banana and the
island state (the first is 4U~1636-53, see \cite{wijnands97}).

The increase in QPO {\it rms} amplitude with photon energy in our
observations shows that the QPO emission is harder than the average
spectrum.  A positive correlation between {\it rms} amplitude and
photon energy in the range from 2 keV to more than 11 keV has also
been observed from kilo-hertz QPOs in other sources (for example, Ford
et al. 1997b and Zhang, W. et al. 1996).  This suggests that all these
QPOs are of similar origin.

Our QPO observations can be compared to the observations of 4U 1608-52
made in early March (Berger et al. 1996).  In all observations only a
single QPO with centriod frequency above 200 Hz has been detected and the QPO 
{\it rms} amplitude increases with photon energy.  The {\it rms} amplitude versus 
energy curve for March 15 and 22 is consistent with that presented for March 3 
(Berger et al. 1996). However, on March 3 and 6, the QPO was with a FWHM of 
5--15~Hz in the PDS averaged over 100 seconds; while on March 9, 15, and 22, 
the QPO can not be tracked as the QPO observed on March 3 and 6, and was with 
a FWHM of 80--140~Hz, which are derived from the average PDS 
over a few thousand seconds. On March 3, the X-ray intensity
was high and the QPO frequency varied in the range 830-890~Hz with no
correlation with X-ray intensity (\cite{berger96}). Comparing individual orbits 
on March 15 and 22, when the X-ray intensity is lower, the QPO frequency
is lower (570--800 Hz) and appears to be positively correlated with
X-ray intensity.  This correlation does not hold when data from
different days are compared.  A similar trend of correlation over one
day time scales and lack of correlation on longer timescales has been
observed for QPO frequency versus X-ray intensity in other X-ray burster 
sources (\cite{ford97a}; \cite{wzhang97}).  However, the correlation of QPO 
frequency with the flux of a blackbody component of the X-ray spectrum was found
to be robust over several months in 4U~0614+091 (\cite{ford97b}).

Kaaret et al. (1997) have interpreted the lack of correlation between
QPO frequency and X-ray intensity of 4U~1608-52 on March 3 as reported in 
Berger et al. (1996) as evidence that the accretion disk is terminated near the 
marginally stable orbit when the source is at high X-ray intensities and high 
mass accretion rates.  The March 15 and 22 observations presented here suggest 
that the QPO frequency may be  correlated with X-ray intensity, at least over
one day time scales, when 4U~1608-52 is at low X-ray intensities.
This is consistent with the interpretation of Kaaret et al. (1997)
since, when the mass accretion rate is low, the disk should be
disrupted by the neutron star magnetosphere or radiation forces
outside the marginally stable orbit and the QPO frequency should then
be correlated with mass accretion rate (Alpar \& Shaham 1985; Miller
et al. 1997).  It is important that additional observations of
4U~1608-52 be obtained over a wide range of X-ray intensities.

It is also possible that the QPOs observed on March 15 and 22
may correspond to a pair of QPOs, 
but only one of the two is detectable in the individual orbit (see Fig.6). 
This would imply that the frequency separation is larger than 233$\pm$22 Hz 
(Fig.6). It is thus reasonable to assume that the neutron spin frequency is 
larger than 233 Hz, within the framework of the beat frequency model discussed 
above. If we interpret the 20 Hz QPO observed on March 15 as due 
to the Lense-Thirring precession in the model of Stella and Vietri (1997), the 
inferred neutron star spin frequency for 4U~1608-52 is between 300 Hz and 365 Hz, 
depending on the tilt angle off the equatorial plane. However, future 
simultaneous detection of all three QPOs (the Lense-Thirring precession 
frequency, the beat frequency and the Kerplerian frequency, as of 20 Hz, 
435-500 Hz and 800 Hz respectively in the case of March 15 observation) 
in 4U1608-52 or a detection of QPO in the X-ray bursts from 4U~1608-52 is 
needed to confirm the above inference.
\begin{acknowledgments}
We appreciate the assistance of RXTE/GOF in data reduction and 
analysis and Dr. S. Dieters of UAH for help with the RXTE data analysis software 
and various suggestions. WY would like to acknowledge partial support from the 
National Natural Science Foundation of China. 
\end{acknowledgments}

\newpage
\begin{figure*}
\caption{Long-term X-ray monitoring of 4U 1608-52 by RXTE/ASM. Our
observations with RXTE were on March 15, 18, and 22. \label{fig1}}
\end{figure*}

\begin{figure*}
\caption{Color-color diagram of 4U 1608-52. Data from March 15, 18 and
22 pointings are plotted as triangles, squares, and circles, respectively.
Each point represent a 240s of observation. The three data points on the
left side of the plot show typical error bars for a hypothetical EXOSAT
ME observation of 4U 1608-52 on the three days with the same energy bands
used in the PCA and the computed ratio of the effective areas of the two
instruments. On March 18, no significant kilo-hertz QPO is detected. 
\label{fig2}}
\end{figure*}

\begin{figure*}
\caption{Average PDS on March 15, 18 and 22 obtained from the 
background-subtracted {\it Event Mode} data. The average level of the 
white noise caused by counting statistics was subtracted. The fits of 
the PDS with the model described in Hasinger and van der Klis (1989) 
are also shown in the plot, with
reduced ${\chi}^{2}$s (38 degrees of freedom) of 1.80, 1.84 and 1.08 
respectively. The rms of HFN was about 13\% in the above PDS. \label{fig3}}
\end{figure*}

\begin{figure*}
\caption{Average Power Density Spectrum in the third orbit on March 15
obtained from the 1-30 keV {\it Event Mode} data. High frequency noise
(HFN) is also visible below 100 Hz. We fit the PDS in 200-2000 Hz range
with a model (solid curve) composed of a linear component plus a Lorentzian
peak at 744 Hz. \label{fig4}}
\end{figure*}

\begin{figure*}
\caption{Average fractional {\it rms} amplitude of QPO as a function
of photon energy. We average the results obtained from {\it Event Mode}
data analysis of the 6 orbits. \label{fig5}}
\end{figure*}

\begin{figure*}
\caption{Average count rates (1-60 keV) vs. QPO frequency in 4U 1608-52
during the low state observation. Data points on March 15 and 22 were
marked as squares and circles. No QPO peaks were observed on March 18. 
\label{fig6}}
\end{figure*}

\newpage
\begin{deluxetable}{lrrrrrr}
\tablewidth{0pc}
\tablecaption{ March 1996 RXTE Observations of 4U 1608-52\tablenotemark{a}}
\tablehead{
\colhead{Orbit}         &\colhead{Duration}     &
\colhead{{Count Rate}\tablenotemark{b}}&
\colhead{{Count Rate}\tablenotemark{c}}&
\colhead{${\nu}_{qpo}$\tablenotemark{d}}&\colhead{FWHM\tablenotemark{e}}        
&
\colhead{{\it rms}\tablenotemark{f}} \\
\colhead{{Start Time}\tablenotemark{g} }                &\colhead{(s)}    &
\colhead{(cps)}& \colhead{(cps)}&
\colhead{(Hz)}&\colhead{(Hz)}       &
\colhead{$(\%)$}}
\scriptsize
\startdata
March 3 &&&2910-3400&830-890&5-15&$\sim6-8$ \nl
March 6 &&&1920-2500&830-870&5-15&$\sim14$ \nl
March 9 &&&610-730&$691\pm^{6}_{6}$&$131\pm^{19}_{19}$&$\sim13.9$ \nl
March 12&&&460-710&&& \nl
March 15 
19:28&1800&300-350&150-180&$800\pm^{9}_{10}$&$78\pm^{25}_{20}$&$13.3\pm^{3.1}_{3
.
4}$ \nl
March 15 21:39&1200&300-340&155-185&&& \nl
March 15 
22:49&2700&280-330&150-175&$744\pm^{12}_{12}$&$116\pm^{40}_{30}$&$14.4\pm^{3.3}_
{
3.7}$ \nl
March 18 19:46&2400&185-225&105-135&&& \nl
March 18 21:10&3000&190-230&105-135&&& \nl
March 22 15:11&1600&415-450&210-250&&& \nl
March 22 
16:14\tablenotemark{h}&3600&415-450&210-240&$638\pm^{11}_{11}$&$126\pm^{25}_{32}
$
&$15.9\pm^{2.9}_{3.1}$ \nl
March 22 
17:50&3600&375-430&200-230&$637\pm^{16}_{17}$&$173\pm^{53}_{41}$&$17.3\pm^{3.8}_
{
3.9}$ \nl
March 22 
19:35&3000&370-430&190-240&$622\pm^{31}_{29}$&$180\pm^{123}_{100}$&$13.6\pm^{5.8
}
_{8.2}$ \nl
March 22 
21:22&2400&370-415&190-230&$567\pm^{21}_{18}$&$134\pm^{93}_{82}$&$14.0\pm^{6.3}_
{
9.3}$ \nl
March 22 23:04&1400&345-415&184-230&&& \nl
\tablecaption{\em Table 2. Observations of QPO in 4U 1608-52 at low intensity 
state.}
\enddata
\tablenotetext{a}{Results before March 15 are from Berger et al. 1996 and
the corresponding count rates are from van der Klis 1997a.}
\tablenotetext{b}{The count rates in our observations were obtained from 1-60 
keV light 
curve with
16s time resolution. The count rates obtained from 3 PCUs on March 22 have
been rescaled to count rates as observed from 5 PCUs. }
\tablenotetext{c}{The count rates in our observations were obtained from 5-60 
keV light 
curve with
16s time resolution and rescaled as from 5 PCUs.}
\tablenotetext{d}{The QPO centroid frequencies were obtained by 
fitting the peak with a Lorentzian.}
\tablenotetext{e}{The FWHM of the Lorentzian peak in PDS.}
\tablenotetext{f}{All has been corrected to fractional {\it rms} amplitude of 
source
intensity in 1-30 keV in our observation. Correction of binning effect has been 
applied.
}
\tablenotetext{g}{Start time (UT) of each orbit (date,hour:minute). }
\tablenotetext{h}{We exclude 50s data of each burst to calculate the PDS in the
orbit.}
\end{deluxetable}

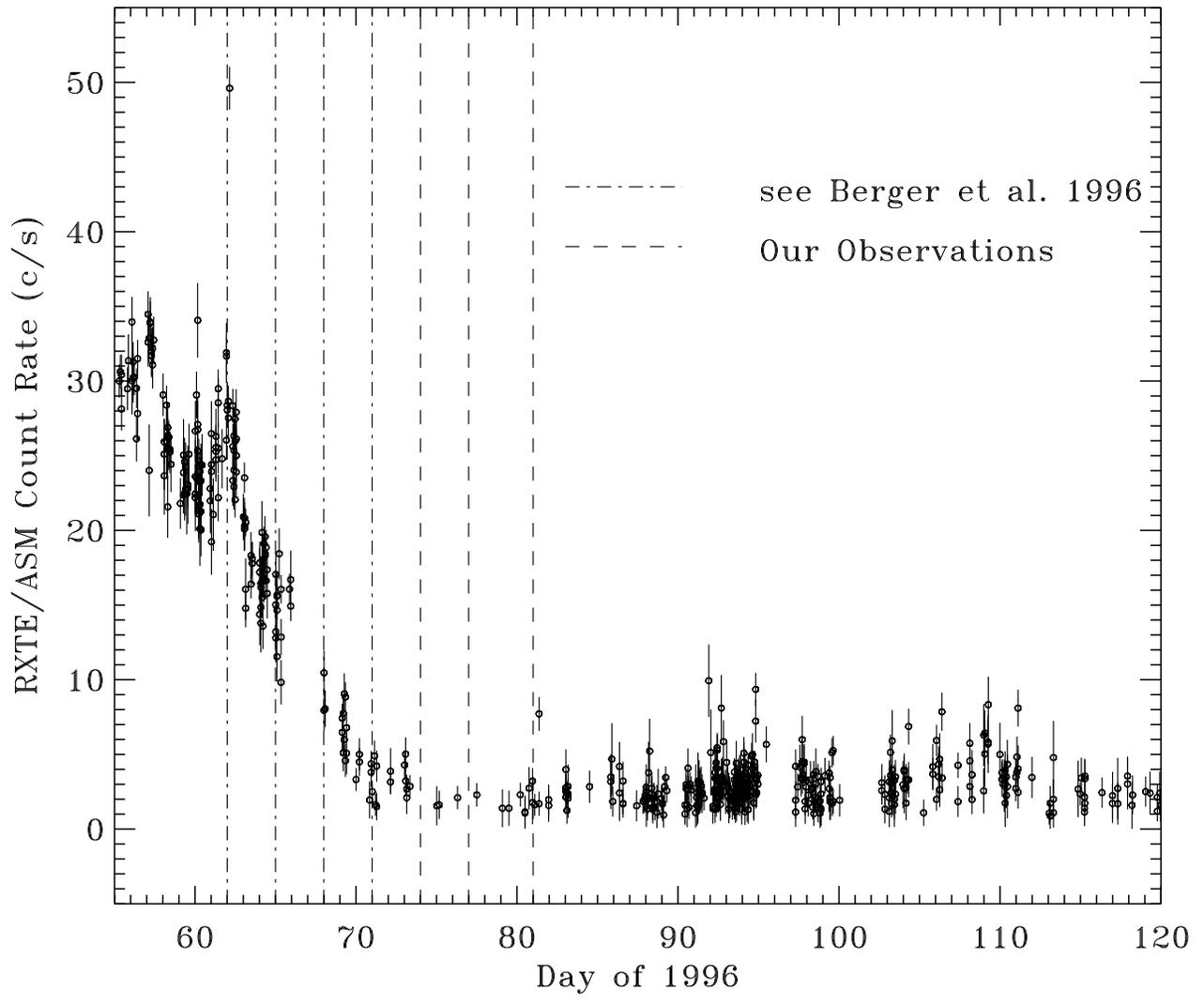
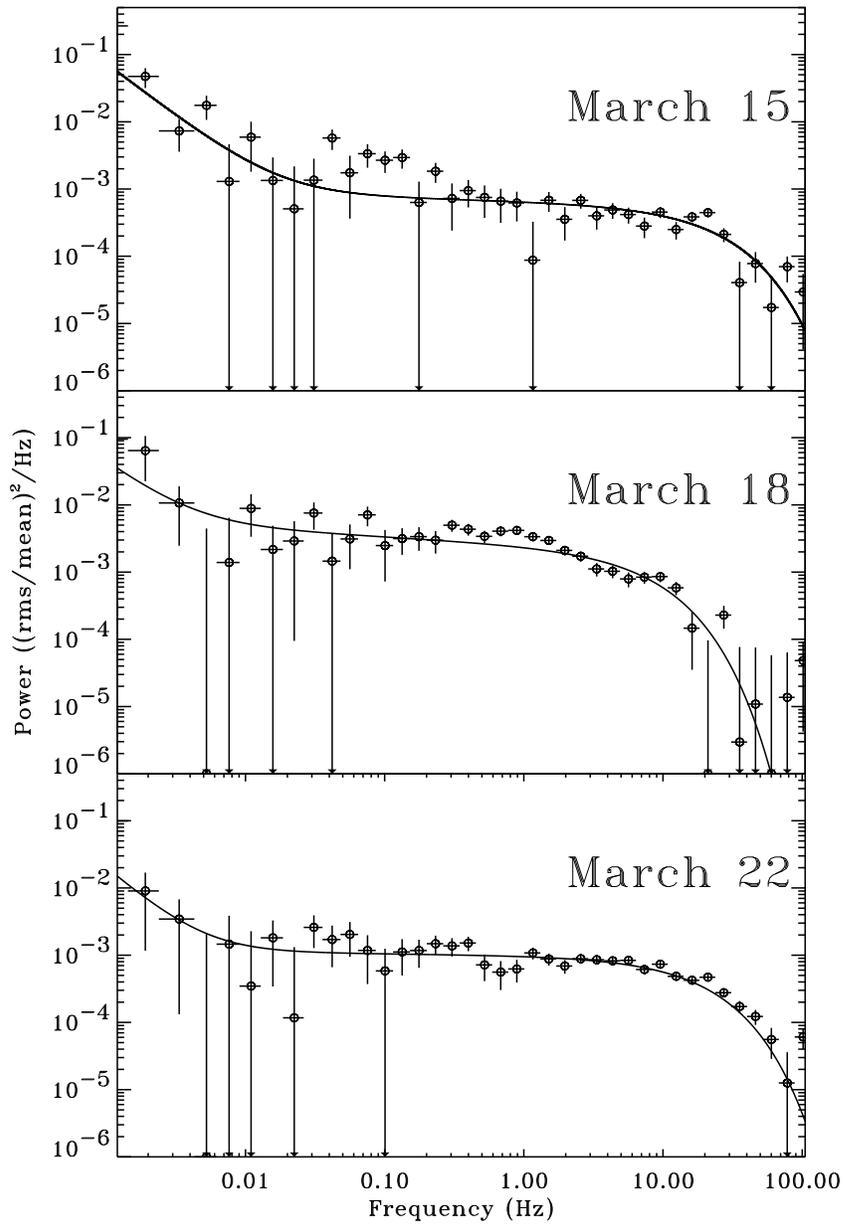
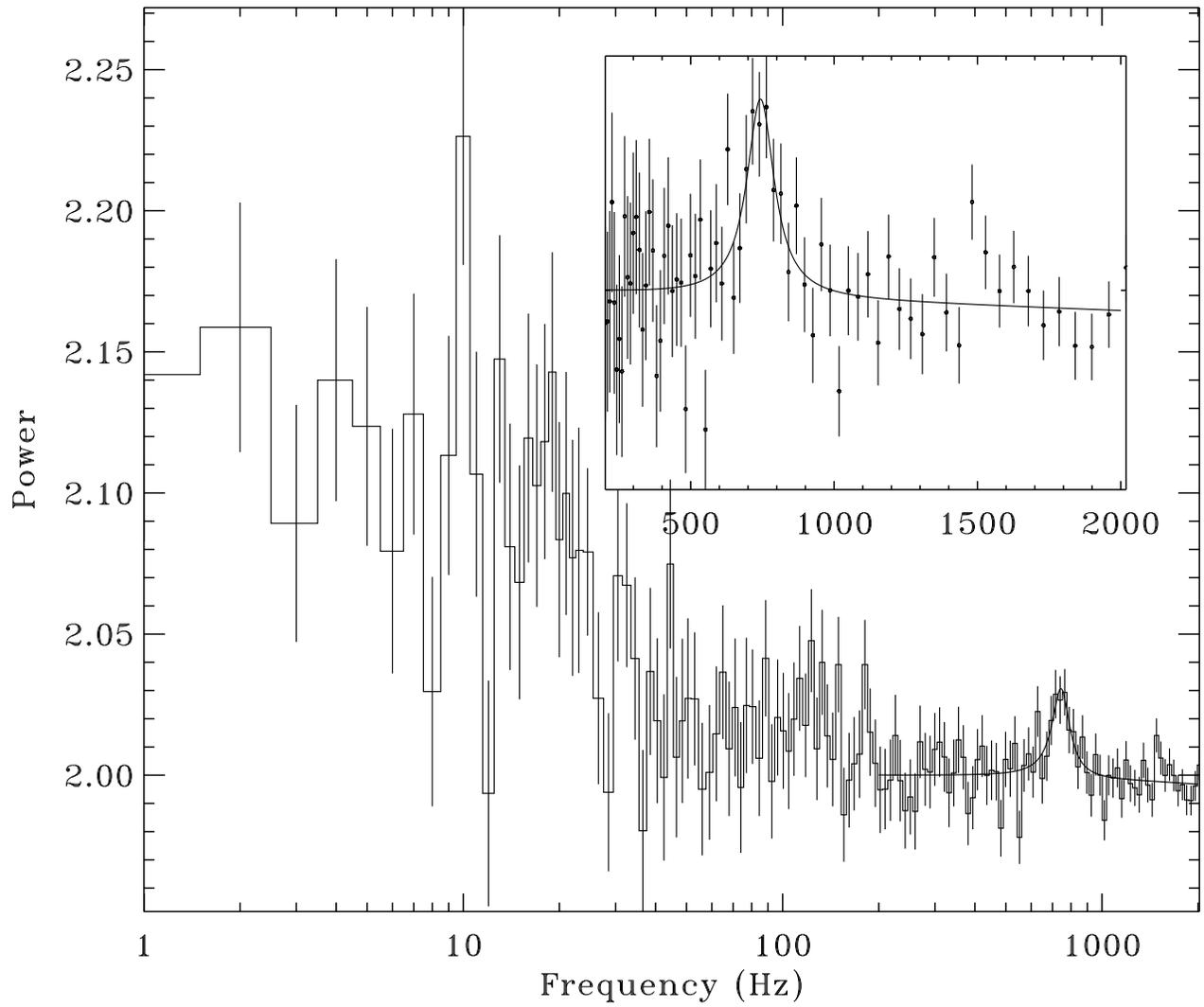
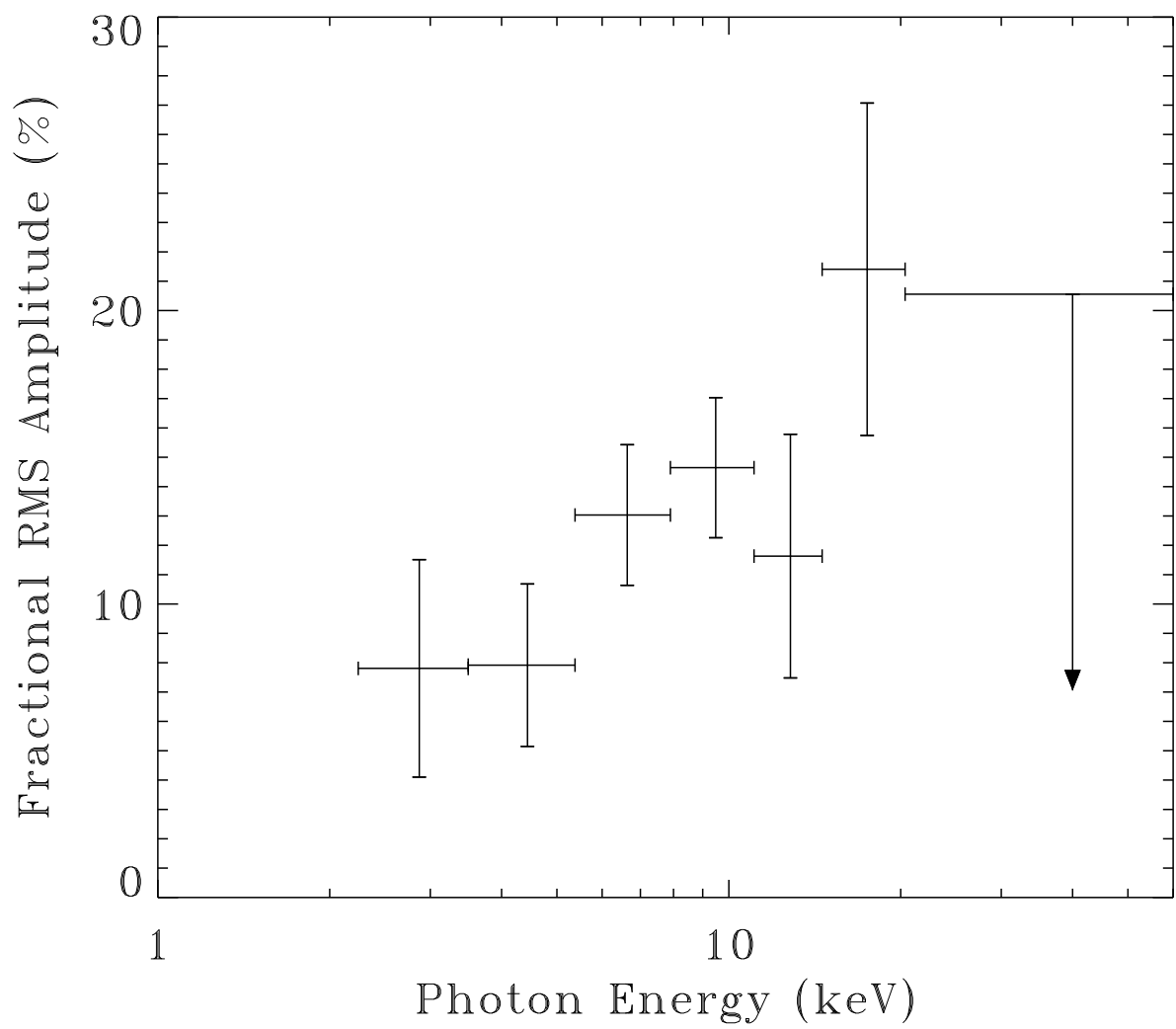
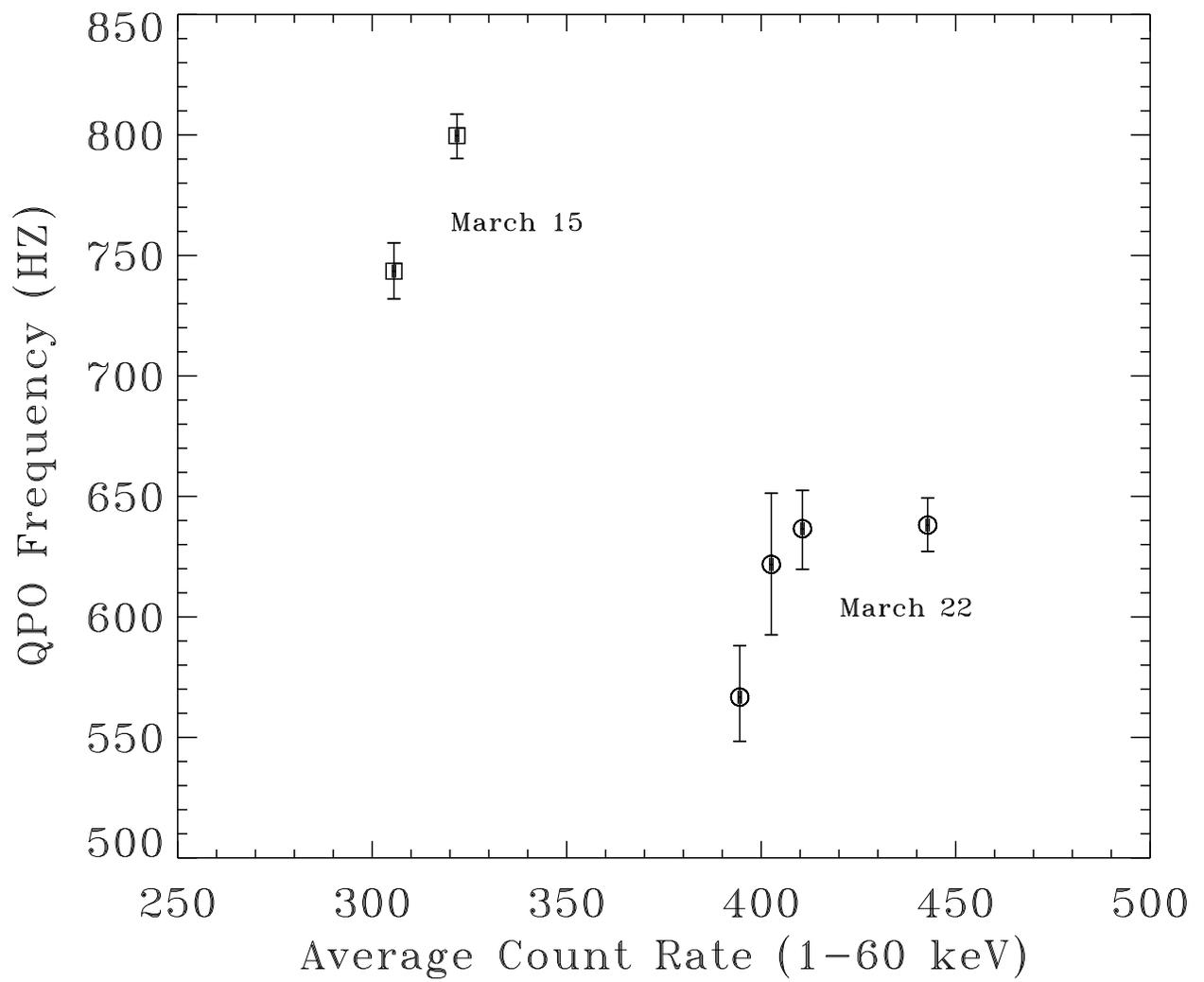
\end{document}